\documentclass[12pt,preprint]{aastex}
 
\def\alwaysmath#1{\ifmmode{#1}\else{$#1$}\fi}

\def\arcsec{\hbox{$^{\prime\prime}$}} 
\def\etal{{et al.~}}

\newcommand\HST{{\sl HST}}

\def\ltsima{$\; \buildrel < \over \sim \;$} 
\def\gtsima{$\; \buildrel > \over \sim \;$} 
\def\lsim{\lower.5ex\hbox{\ltsima}} 
\def\gsim{\lower.5ex\hbox{\gtsima}} 
\def\rhalf{\alwaysmath{r_{1/2}^{\scriptscriptstyle\rm BSS}}} 
\def\rcore{\alwaysmath{r_{c}}} 
\def\frgb{\alwaysmath{f_{\scriptscriptstyle\rm RGB}}} 
\def\bsshb{\alwaysmath{F_{\scriptscriptstyle\rm HB}^{\scriptscriptstyle 
   \rm BSS}}} 
\def\bbsshb{\alwaysmath{F_{\scriptscriptstyle\rm HB}^{\scriptscriptstyle 
   \rm bBSS}}}

\def\arcsec{\hbox{$^{\prime\prime}$}} 

\lefthead{Ferraro, \etal} 
\righthead{Blue Straggler Stars in Galactic Globular Clusters} 
 
\slugcomment{Submitted to the Astrophysical Journal} 
 
\begin{document} 
 
\title{Blue Straggler Stars: a direct comparison of Star counts and 
population ratios in  six Galactic Globular Clusters\footnote{Based on 
observations with the NASA/ESA {\it Hubble Space Telescope}, obtained at 
the Space Telescope Science Institute, which is operated by AURA, Inc., 
under NASA contract NAS5-26555}}

\author{Francesco R. Ferraro\altaffilmark{2}, 
	Alison Sills\altaffilmark{3}, 
	Robert T. Rood\altaffilmark{4}, 
	Barbara Paltrinieri\altaffilmark{5}, 
	Roberto Buonanno\altaffilmark{5,6} 
        }

\altaffiltext{2}{Dipartimento di Astronomia, Universit\`a
di Bologna, via Ranzani 1, 
I--40126 Bologna, Italy; ferraro@apache.bo.astro.it} 
\altaffiltext{3}{Department of Physics and Astronomy, McMaster University, 1280 Main Street West, Hamilton, ON, L8S 4M1, Canada; asills@mcmaster.ca} 
\altaffiltext{4}{Astronomy Dept., University of Virginia, Charlottesville,  
VA 22903-0818, USA; rtr@virginia.edu } 
\altaffiltext{5}{Osservatorio Astronomico di Roma, Via Frascati 33, 00040
Monte Porzio Catone, Italy}  
\altaffiltext{6}{Universit\`a di Roma "Tor 
Vergata", Dip di Fisica, Via della Ricerca Scientifica, I-00133 Roma, Italy}
\begin{abstract} 
 
The central regions of six Galactic Globular Clusters (GGCs) (M3, M80,
M10, M13, M92 and NGC 288) have been imaged using {\it HST-WFPC2} and
the ultraviolet (UV) filters (F255W, F336W).  The selected sample
covers a large range in both central density ($\log \rho_0$) and
metallicity (${\rm [Fe/H]}$).  In this paper, we present a direct
cluster-to-cluster comparison of the Blue Stragglers Stars (BSS)
population as selected from $(m_{255},m_{255}-m_{336})$ Color
Magnitude Diagrams (CMDs).  We have found: {\it(a)} BSS in three of
the clusters (M3, M80, M92) are much more concentrated toward the
center of the cluster than the red giants; because of the smaller BSS
samples for the other clusters we can only note that the BSS radial
distributions are consistent with central concentration; {\it (b)} the
specific frequency of BSS varies greatly from cluster to cluster.  The
most interesting result is that the two clusters with largest BSS
specific frequency are at the central density extremes of our sample:
NGC 288 (lowest central density) and M80 (highest). This evidence
together with the comparison with theoretical collisional models
suggests that both stellar interactions in high density cluster cores
and at least one other alternate channel operating low density GGCs
play an important role in the production of BSS. We also note a
possible connection between HB morphology and blue straggler
luminosity functions in these six clusters.
\end{abstract} 
 
 
\section{Introduction}  
\label{sec:intro} 
 
Stellar collisions are thought to be the dominant channels for 
the formation and evolution of various types of unusual stellar 
objects and binary systems in the dense stellar environments, such as the 
central regions of Galactic Globular Clusters (GGCs). Among the 
possible {\it collisionally induced} populations, Blue Stragglers 
Stars (BSS) are surely the most studied.  BSS were first discovered by 
Sandage (1953) in M3. In recent years the realization that BSS are the 
ideal diagnostic for a quantitative evaluation of the effects of 
dynamical interactions inside star clusters has led to a remarkable 
burst of activity in the search for and the systematic study of BSS in 
clusters. In addition, the Hubble Space Telescope (\HST), with its 
high angular resolution and UV imaging capabilities, has made it 
possible to search for candidates BSS in the cores of highly 
concentrated GGCs. 
 
Within this framework we are performing an extensive {\HST}-UV survey in the 
cores of a selected sample of GGCs searching for various products of 
binary evolution.  In particular, we are constructing an homogeneous 
database of BSS in the UV-bands. Some results have been already 
published in a series of papers presenting the BSS content in 
individual clusters (see Ferraro \etal 1997a, 1999a,  
2001 and Paltrinieri \etal 1998, Bellazzini \etal 2002, Ferraro, 
Paltrinieri \& Cacciari, 1999, for a review). In this series of papers 
we empirically demonstrated that UV-CMDs are the most 
appropriate planes for the study of BSS. 
 
In this paper, we present a direct comparison of the BSS content in 
six intermediate to high density ($\log \rho_0\sim 2$--5.4) GGCs with 
intermediate to low metallicity: M13 (NGC 6205), M10 (NGC 6254), M80 
(NGC 6093), M3 (NGC 5272), M92(NGC 6341) and NGC 288. In 
\S \ref{sec:res} we present the BSS samples discovered in each 
cluster on the basis of the UV-CMDs, and we discuss the properties and the 
radial distributions of BSS with respect to normal cluster 
stars. In \S \ref{sec:teo} we present the comparison of the 
BSS distribution in the UV-CMDs with that expected from theoretical 
collisional models.

\section{Observations and  data analysis} 
\label{sec:obs} 
   
The data-set used in this work consist of a series of {\HST}-WFPC2
exposures covering the central regions of six GGCs: M3, M10, M13, M80,
M92 and NGC 288.  In Table 1 we summarized the main characteristics of
each cluster: metallicity (${\rm [Fe/H]}$) from Zinn \& West (1984),
central density ($\rho_0$ and mass (from Trager \& Meylan 1993),
distance ($d$) (from Ferraro \etal 1999b), and central dispersion
velocity ($\sigma_0$) (from Trager \& Meylan 1993).  For each cluster,
the {\it Planetary Camera} (PC) was roughly centered 
on the cluster center (in the case of NGC
288 the cluster center was outside the field of view of the PC - see
Figure 1 by Bellazzini \etal 2002). 
By adopting this observational strategy a significant fraction 
of the cluster luminosity has been sampled with a single HST pointing,
In Table 3 (column 2) the fraction of the cluster luminosity sampled 
($L_S/L_T$) in each cluster is reported.
 Here we present the results
obtained using the mid-UV (F255W) and near-UV (F336W) filters, those
best suited for picking out BSS.  Table 2 lists the total duration of
the exposures in each filter for the six clusters.
   
All the photometric reductions (with the exception of NGC 288; see
Bellazzini \etal 2002) have been carried out using ROMAFOT (Buonanno
\etal 1983), a package specifically developed to perform accurate
photometry in crowded fields.  The standard procedure described in
Ferraro \etal (1997a) was adopted.  More details on the reduction
procedure and the entire dataset secured in these clusters can be
found in some specific papers already published (see for example
Ferraro \etal 1997a for M3 and Ferraro \etal 1999a for M80).  Briefly,
the images in the same filter were aligned and combined in order to
obtain a median image in each filter.  We used the combined F255W
image as reference frame for searching the objects in each field.
Then the PSF fitting procedure was separately performed on each
individual frame, and an average magnitude was computed for each star.
The instrumental magnitudes were finally transformed into the STMAG
system, using Table 9 of Holtzmann \etal (1995).

\section{Results}  
\label{sec:res} 
 
In the UV cluster light is dominated by hot stars, specifically the
blue HB stars and the BSS.  (see for example Figure 1 by Ferraro,
Paltrinieri \& Cacciari 1999).  Since the BSS are one of the hottest
sub-populations in the cluster, they are easily separated from the
cooler Turn-Off and SGB stars. Indeed, in previous papers (Ferraro
\etal 1997a, 1999a, 2001 and Paltrinieri \etal 1998,
Ferraro, Paltrinieri \& Cacciari, 1999) we have shown that UV-CMDs, in
particular the $( m_{255},~ m_{255}-m_{336})$ plane, are ideal for
selecting BSS.
 
Figure 1 shows the ($m_{255},~m_{255}-m_{336}$) CMDs for the six
clusters.  More than 50,000 stars are plotted in the six panels of
Figure 1.  As can be seen, in this plane, the BSS populations define a
clean and well-defined sequence spanning $\sim 3$ mag in
$m_{255}$. Further, they are clearly distinguishable from SGB-TO
stars.  In order to select the BSS samples we followed the same
criteria that we adopted for M3. In Ferraro \etal (1997a), the global
population of BSS was divided in two sub-samples: (a) {\it bright} BSS 
with $m_{255}<19$ and (b) {\it faint} BSS with $19.0<m_{255}<19.4$.
To apply the same criteria to the other clusters shown in Figure 1,
the CMD of each cluster has been shifted to match that of M3  by using
the brightest portion of the HB as the normalization
region. 
Figure 1 shows the UV-CMD after the alignment.  Thus, the solid
horizontal line (at $m_{255}=19$) in the figure shows the cut-off
magnitude for the bBSS sample.  In this paper we are
comparing the BSS content in clusters with quite different structural
parameters. The level of difficulty in the detecting
and the measuring of (especially faint) BSS varies from cluster to
cluster. Thus, we decided to compare the
samples by using only the {\em bright BSS samples} (hereafter bBSS).
The number of
bright BSS ($N_{bBSS}$) identified in each cluster is listed in Table 3.

\subsection{bBSS radial distribution} 
\label{subsec:rad} 
 
As pointed out by many authors most BSS in GGCs are found to be 
centrally concentrated with respect to {\em normal} stars.  Since the 
central relaxation time in these systems is much less than the cluster 
age, this result is generally ascribed to dynamical mass segregation 
and can be interpreted as an evidence that BSS are more massive than 
the comparison stars. 
 
We have determined the radial distributions of the bBSS of the six
GGCs using the RGB stars (from the tip down to the SGB ) as a
``reference'' population. The RGB stars have been selected in the
$(m_{555}, m_{336}-m_{555})$ CMD, i.e. roughly a $(V,~U-V)$ CMD in order
to reduce any bias which might be introduced by the poor photometry
for the very red stars in the UV bands.  The cumulative radial
distributions for the bBSS and the RGB stars are plotted in Figure 2
as a function of the projected distance ($r$) from the cluster center.
It is evident from the plots that the BSS (solid line) are
significantly more centrally concentrated than RGB stars (dotted line)
in M3, M92 and M80. A Kolmogorov-Smirnov test has been applied to the
two distributions, in each cluster, to check the statistical
significance of the detected differences.  The test yields the
probability that the bBSS population and the RGB stars are
extracted from the same {\it parent} distribution. The probability
values obtained in each cluster are reported in Figure 2. The effect
is strongly statistically significant in the case of M3, M92 and
M80. For NGC 288 and M10 the BSS are more concentrated that the RGB
but the results are significant at only roughly the $2\sigma$ and
$1\sigma$ levels respectively.

At first glance the radial distribution of the BSS in M13 appears to
be indistinguishable from the other cluster stars. Since it is
difficult to imagine any BSS formation scenario in which the BSS are
not at least somewhat centrally condensed, we have more carefully
examined the role of the small M13 bBSS sample. A series of Monte
Carlo simulations have been performed in which $10^3$ subsamples of
the same number of the bBSS population found in M13 (16) have been
randomly extracted from the M3 bBSS population (72). Then, the radial
distribution of each subsample was compared with that of the RGB stars
of M3. A KS test has been applied to quantify the significance of the
differences. In these simulations drawn from a sample in which the
central concentration of the BSS has been convinvingly shown, 40\% of
the M13-like samples have $p>20\%$ that the BSS and RGB have the same
distribution. Less than 10\% have as a convincing a case for different
distributions as in M3. About $6\%$ of the M13-like samples have $p$
as large as for the true M13 comparison. In another test, two M13 BSS
were arbitrary moved from the middle of the distribution to the inner
part and $p$ dropped from 0.77 to 0.34. We conclude that the M13 bBSS
sample is too small to make a definitive statement about its central
concentration. Our M13 results are consistent with no central
concentration, but they are not inconsistent with a central
concentration even as large as that of M3.

The
case of NGC 288 deserves a short comment since the location of the
center in this cluster is quite uncertain.  The radial distribution
plotted above could be affected by large uncertanties. In accordance
with Bellazzini \etal (2002) we adopted the coordinates listed by
Webbink (1985), and the cluster center turns out to be located just
outside the field-of-view of the PC (see their Figure 1).
 
We can make an additional quantitative comparison by using the
parameters \rhalf\ (the radius containing half the bBSS sample) and
\frgb (the fraction of the total RGB sample contained within
\rhalf). Values for \rhalf\ and \frgb\ are listed in Table 3 along
with the core radius (\rcore). With the exception of NGC 288 (for
which we took the value from Trager, Djorgovski \&
King, 1993), the \rcore\ was independently determined in each
cluster by fitting a King model to the observed density profile (see
for example Figure 4 in Ferraro \etal 1999a). These values confirm the
general impression from Figure 2,that the BSS are much more
concentrated than the RGB in M80, M3 and M92. Thus, even considering
the brightest part of the BSS distribution in M80, we have fully confirmed
the previous results by Ferraro \etal 1999a, who suggested that the BSS
in M80 are the most concentrated population observed in a GGCs.

In Figure 3 the magnitude distributions (equivalent to a luminosity
function---LF) of bBSS for the six clusters are compared.  In doing
this we use the parameter $\delta m_{255}$ defined as the magnitude of
each bBSS (after the alignment showed in Figure 1) with respect to the
magnitude threshold (assumed at $m_{255}=19$ - see Figure 1). Then
$\delta m_{255}= m^{bBSS}_{255}-19.0$. From the comparison shown in
Figure 3 ({\it panel(a)}) the bBSS magnitude distributions for M3 and
M92 appear to be quite similar and both significantly different from
those obtained in the other clusters. This is essentially because in
both these two clusters the bBSS magnitude distribution seems to have
a tail extending to brighter magnitudes (the bBSS magnitude tip
reaches $\delta m_{255}\sim -2.5$). A KS test applied to these two
distributions yields a probability of $93\%$ that they are extracted
from the same distribution. In {\it panel(b)} we see that the bBSS
magnitude distribution of M13, M10 and M80 are essentially
indistinguishable from each other and significantly different from M3
and M92. A KS test applied to the three LFs confirms that they
are extracted from the same parent distribution. Moreover, a KS test
applied to the total LFs obtained by combining the data for the two
groups: M3 and M92 ({\it group(a)}), and M13, M80 and M10 ({\it
group(b)}) shows that the the bBSS-LFs of {\it group(a)} and 
{\it group(b)} are
not compatible (at $3\sigma$ level).

It is interesting to note that the clusters grouped on the basis of
bBSS-LFs have some similarities in their HB morphology. The three
clusters of {\it group(b)} have an extended HB blue tail; the two clusters
of {\it group(a)} have no HB extention. Could there be a connection between
the bBSS photometric properties and the HB morphology? This
possibility needs to be further investigated.

Finally we note that the magnitude function for the bBSS in NGC 288
(solid line in {\it panel (b)}) is significantly different than that of
all the other clusters.

\subsection{Specific Frequency} 
\label{subsec:fre} 
 
In order to properly compare the BSS populations in different 
clusters, the BSS number must be normalized to account for the size 
of the total cluster population. For this reason we have defined 
various specific frequencies. In Ferraro, Fusi Pecci \& 
Bellazzini (1995) we used $S4_{\scriptscriptstyle\rm BSS}$ (the BSS 
number normalized to the integrated bolometric luminosity of the 
surveyed region in units of $10^4 L_{\odot}$).  Analogously, in 
Ferraro \etal (1999a) we defined a more appropriate specific frequency: 
 
$$ \bsshb = {{N_{\scriptscriptstyle\rm BSS}}  
\over {N_{\scriptscriptstyle\rm HB}}} $$ 
 
\noindent where $N_{\scriptscriptstyle\rm BSS}$ is the number of BSS and 
$N_{\scriptscriptstyle\rm HB}$ is the number of HB stars in the same 
area. 
The two ratios are similar, since in 
absence of any special segregation of HB stars
with respect to the "normal" cluster-star, the observed number of HB
stars $N_{\scriptscriptstyle\rm HB}$ is an excellent indicator of
 the sampled cluster luminosity. In fact, the simple relation by Renzini
 \& Buzzoni   (1986):  
 $$ N_{j} =B(t) L_S t_j$$
 
\noindent easily links the number of stars sampled 
 in any post main sequence stage ($ N_{j}$) with the 
 sampled cluster luminosity ($ L_S$) and the duration of the phase ($t_j$).
 $ B(t)$ is the specific flux of the population. Following 
 Renzini \& Buzzoni (1986) we assume
 $ B(t) \sim 2\times 10^{-11} {~\rm stars}\, L_{\odot}^{-1}\, {\rm yr}^{-1}$ for an old
 population of $10^{10}\,{\rm yr}$.
 
 Thus, assuming $t_{\rm HB}\sim 10^8\, {\rm yr}$, we can expect to sample
 $2\times 10^{-3}$ HB stars for each solar luminosity sampled 
 in a given stellar population.

The $\bsshb$ ratio has the advantage that  it is a purely
observative quantity and it can
be easily computed in the 
UV-CMDs since the HB population is quite bright in these planes and 
the HB sequence is well separated from the other branches. 
However, while in the following discussion we will mainly 
use the $\bsshb$ ratio,  the $S4_{\scriptscriptstyle\rm BSS}$ 
is also computed for each cluster 
 and the values are listed in Table 3 for sake of comparison.

As discussed in Section 3 we consider only the bright BSS, hence 
$N_{\scriptscriptstyle\rm bBSS}$ is the number of bright bBSS 
(according to the original definition by Ferraro \etal 
(1997a)). The values of \bbsshb\ are listed in Table 3.  As can be seen 
\bbsshb\ varies by a factor 13 from M13 to NGC 288. Interesting enough 
NGC 288, the cluster with the lowest central density, shows the
largest bBSS specific frequency. As already noted by Bellazzini \etal
(2002) this cluster turns out to have as many bBSS as HB stars over
the sampled area. Though the number of bBSS detected in the cluster is
intrinsically small ($\sim 30$) the result is surprising: \bbsshb\ is
comparable with what found in the central region of the
core-collapsing cluster M80 by Ferraro \etal (1999a). If we consider
only the PC field of view, the \bbsshb\ in M80 rises up to 0.7.

The central density of a cluster is not the only factor which
determines the number of collision products we expect to
see. Collisions involving binary star systems are more likely than
collisions between single stars and have a significant probability of
producing a stellar merger. We used equation 14 from Leonard
(1989) in order to estimate the number of binary-binary ({\it bb})
encounters occuring per Gyr ($N_{bb}$) in the core of each cluster
listed in Table 1.  Adopting the distance modulus listed in Table 1,
the $r_c$ listed in Table 3, $\sigma_0$ and the central density
$Log(\rho_0)$ listed in Table 1 (from Pryor \& Meylan (1993)),
assuming an average mass of 0.2$M_{\odot}$ (Kroupa 2001),
 we computed
$t_{bb}$ (the mean interval between collisions) from equation (14) by
Leonard (1989) and derived the expected number of encounters, per Gyr,
as a function of the binary fraction ($f_{b}$) in the core.

In doing this, the semimajor axis which separates hard from soft
binaries ($a_{hs}$) has been computed for each cluster according
to the following relation:
$$ a_{hs} = {{G M} \over {9 \sigma_0^2}}$$
derived by combining equation 4 of Leonard \& Linnell (1992)
with the Kepler's third law.
The semimajor axes computed for each cluster and expressed
in AU are listed in 
Table 4.

 Following
Leonard (1989) we multiplied $t_{bb}$ by a factor of two in order to
take into account the fact that not all encounters lead to a physical
collision.  The rate of single-binary collisions can also be estimated
from Leonard's results. In his equation 8 if one assumes that the
pericenter distance is proportional to binary semimajor axis, like in
binary-binary collisions, and that $M_1=M_\star$,
$M_{2}(binary)=1.5\, M_\star$, one finds a factor of 5 instead of his 6
in the cross section.  Thus the number of single-binary encounters is
roughly the 5/6 the number of binary-binary encounters. 

In Table 4 we list the results with three different $f_{b}$ values,
respectively  0.2 and 1.00. These values should be considered as
indications of the effectiveness of physical collisions in the cluster
cores. It is clear that the {\it bb} (or {\it sb}) collision channel
is more than one order of magnitude more efficient in a dense cluster
like M80 than in a low density cluster like NGC 288. Therefore, the
high specific frequency of blue stragglers in NGC 288 suggests that
the binary fraction in NGC 288 is much higher than the binary fraction
in M80. Only then would one expect a similar encounter frequency in
the two clusters. Note that a cluster like M80 may have originally had
a higher binary fraction but because of the efficiency of encounters,
those primordial binaries were ``used up'' early in the history of the
cluster, producing some collisional BSS which 
have evolved away from the MS (see for example the evolved E-BSS
population found in M3 and M80 by Ferraro 1997a, 1999a).

We should note that on the basis of the 
simulation results shown in Figure 8 and in Table 4 the binary-binary
channel can still account for the number of the BSS 
observed in NGC288 if a large enough percentage of binary fraction
is assumed to reside in the core. 
 
However, without invoking {\it ad hoc} binary content, 
{\bf a more natural explanation}
 for the origin of BSS in NGC 288 (as discussed in
Bellazzini \etal (2002)) is the mass transfer process in primordial
binary systems (Carney et al 2001). Here we probably have another
confirmation of the scenario suggested by Fusi Pecci \etal (1992, and
references therein): BSS living in different environments have
different origins.
  
\section{Collisional Models} 
\label{sec:teo} 
 
From our sample of six clusters with rather diverse bBSS populations
we can select pairs with similarities in parameters like density,
metallicity, velocity dispersion, or we can search for trends in bBSS
populations as some parameter varies. There are two quite striking and
unsuspected results. First, the two clusters at the high and low
density extremes, M80 and NGC 288, have the largest bBSS specific
frequencies.  Second, two clusters which are in most ways quite
similar, M3 and M13, have very different bBSS populations.

To aid our understanding of the BSS populations in these clusters, we
present a comparison of the photometric characteristics of the BSS
observed in the selected clusters with some collisional models. The
models we used in this paper are described in detail in Sills \&
Bailyn (1999), and have been applied to 47 Tucanae (Sills \etal 2000).
We assume that the blue stragglers in the central regions of the six
clusters are all formed through stellar collisions between single
stars during an encounter between a single star and a binary system.
While binary-binary collisions may well be important, we do not
currently have the capability of modeling the BSS they might produce.
The trajectories of the stars during the collision are modeled using
the STARLAB software package (McMillan \& Hut 1996). The masses of the
stars involved are chosen randomly from a mass function for the
current cluster and a different mass function which governs the mass
distribution within the binary system.  A binary fraction, and a
distribution of semi-major axes must also be assumed.  The output of
these simulations is the probability that a collision between stars of
specific masses will occur. We have chosen standard values for the
mass functions ($dN(M) \propto m^{-(1+x)}dM$) and binary
distribution. The current mass function has an index $x=-2$, and the
mass distribution within the binary systems are drawn from a Salpeter
mass function ($x=1.35$). We chose a binary fraction of 20\% and a
binary period distribution which is flat in $\log P$. The total
stellar density was taken from the central density of each
cluster. The effect of changing these values is explored in Sills \&
Bailyn (1999).  The collision products are modeled by entropy ordering
of gas from colliding stars (Sills \& Lombardi 1997) and evolved from
these initial conditions using the Yale stellar evolution code YREC
(Guenther \etal 1992). The models reported here used a metallicity of
${\rm [Fe/H]} =-1.6$, which is approximately correct for 4 of the
clusters investigated in this paper. BSS populations should not be
strongly dependent on metallicity.  By weighting the resulting
evolutionary tracks by the probability that the specific collision
will occur, we obtain a predicted distribution of BSS in the
color-magnitude diagram.
 
In order to explore the effects of non-constant BSS formation rates, we 
considered a series of truncated rates.  In these models we assumed 
that the BSS formation rate was constant for some portion of the 
cluster lifetime, and zero otherwise.  This assumption is obviously 
unphysical---the relevant encounter rates would presumably change 
smoothly on timescales comparable to the relaxation time.  However 
these models do demonstrate how the distribution of BSS in the 
color-magnitude diagram depends on when the BSS were created, and 
thus provide a basis for understanding more complicated and realistic 
formation rates. 
 
\subsection{Comparisons with data} 
 
Some comparisons between the models and the HST observations are shown
in figures \ref{fig:distributions}--\ref{fig:clustersB}. In figures
\ref{fig:distributions}, \ref{fig:lumfunc} and \ref{fig:tempfunc}, we
compare our theoretical predictions to the BSS population of M80. The
theoretical predictions have been normalized by the total number of
predicted blue stragglers for each cluster; this allows us to more
easily compare the shapes of the distributions. Figure
\ref{fig:distributions} is an example of the predicted distribution of
BSS in the $m_{255},~m_{255}-m_{336}$ color-magnitude diagram for M80. The
theoretical predictions are shown as greyscale contours, with the darker
colors indicating more BSS in that region of the CMD. The observed
blue stragglers are shown as crosses. Each of the panels is a
different assumption for the BSS formation rate, as indicated on the
top of the panel. The differences between the models can be understood
in terms of lifetimes of the individual collision products which make
up the distributions. For example, if BSS production stopped 5 Gyr
ago (center panel), we predict that there should be no observed bBSS at
present because all the massive BSS would have had time to evolve off
to the giant branch. At the other extreme, if all the BSS were formed
in the last Gyr (panel marked ``1 Gyr to now''), only a few of the very
brightest (i.e. most massive) would have even started moving towards
the subgiant branch. It is worth pointing out that the brightest BSS
shown in these predictions are not on the standard zero age main
sequence, but begin their lives to the red. These are stars which were
formed from the merger of two fairly evolved main sequence stars, and
so they do not have enough hydrogen fuel left at their cores after the
merger to return to the ZAMS (Sills \etal 1997).
 
Figure \ref{fig:lumfunc} gives the luminosity function for each of the
model distributions for M80. The data are shown as the solid line,
while the dotted line is the theoretical prediction for the formation
times shown on the top of each panel. The luminosity functions are
good probes of the mass distribution of blue stragglers, since the
correlation between mass and luminosity is quite tight. The luminosity
functions show, as do the CMD distributions, that bright BSS require
recent formation. They also show that the relative numbers of brighter
and fainter bBSS is not correct in our models. If we model the fainter
bBSS very well (as in the panel marked ``stopped 2 Gyr ago''), then
our models under-predict the number of blue stragglers between
$m_{255}=17$ and 18.
 
Figure \ref{fig:tempfunc} gives the equivalent ``temperature function'' 
for M80. This is the cumulative number of stars in each color 
(i.e. temperature) bin, in direct analogy to the luminosity 
function. The distribution of blue stragglers in color can be used to 
study the age or age spread of the BSS. Evolutionary tracks for stars 
of these masses are confined to a fairly narrow range in luminosity 
between the main sequence and the base of the giant branch, 
so the luminosity function does not give the entire picture. The 
temperature function is useful for distinguishing between populations 
which have lots of main sequence stars and populations with many 
subgiants. As in figure \ref{fig:lumfunc}, the data are shown as the solid 
line, while the dotted line is the theoretical prediction for the 
formation times shown on the top of each box. 
 
We performed KS tests on the luminosity and temperature functions for 
each cluster and each theoretical distribution. Figures 
\ref{fig:clustersA} and \ref{fig:clustersB} gives the best fit model 
distribution for each cluster, its luminosity function, and its 
temperature function. We will comment on each cluster individually 
here, and then draw some general conclusions in the next section. 
 
M80: The fainter bBSS are very well fit by a distribution in which
formation was truncated about 2 Gyr ago. However, there is a
significant population of BSS which extends up to 2.5 magnitudes above
the turnoff, which is not predicted by such a model and which suggests
that some BSS formation continues until the present.  Ferraro \etal
(1999a) suggested that the large BSS population in M80 is due to the
stellar interactions which are delaying the core collapse process. Now
we add the evidence that many of these BSS are relatively old ($t>2$
Gyr).  These results together suggest that the process of holding off
the core collapse can be rather extended.
 
M10: The BSS in this cluster are fit by different 
formation models in the temperature and luminosity planes. However, 
the difference between the KS probability for the best fit to 
luminosity function (BSS formation ended 2 Gyr ago) and the 
best fit for temperature function (BSS formation ended 1 
Gyr ago) is small. The luminosity function has a 16\% probability of 
being drawn from the first distribution, and a 10\% probability of 
being drawn from the second distribution.  
 
M13: Of all the clusters in this paper, this one is the most
believable for showing a truncated BSS distribution. There is no
evidence for very bright BSSs close to the main sequence. The only BSS
brighter than $\delta m_{255}=2.5$ is very red, consistent with an
evolved object. 
    
M3: None of the collisional models fit this cluster very well. The 
best fit for the luminosity function, as shown in figure 
\ref{fig:clustersB}, is a model in which the BSS stopped forming about 
1 Gyr ago (3\% KS probability). However, the temperature function for 
the same model has almost zero probability of being drawn from the 
same distribution as the data, and the best fit temperature function 
is the scenario in which the BSS stopped forming 2 Gyr ago. In all 
cases, including the models which predict BSS formation right up to the 
present day, we under-predict the number of very bright blue 
stragglers. 
 
M92: The best fit to both the luminosity and temperature functions is 
a model in which the BSS formation was truncated 1 Gyr ago. However, 
there are also very bright blue stragglers in this cluster, suggesting 
that blue straggler formation has continued to the present day or that 
the blue stragglers produced in this cluster are not well modeled by 
collisions between two stars. 
 
NGC 288: This cluster is significantly different from the previous 5,
in that we do not expect that the BSS formed in this cluster are the
products of stellar collisions. The density is simply too low. We also
know that the binary fraction in the core is measured to be $\sim 10$
to 38\% (Bellazzini \etal 2002), so BSS formed in this low density
cluster should be the result of binary evolution rather than stellar
collisions.   By assuming that
binary merger BSS are similar to collisional BSS we can still expore
the BSS formation history in NGC~288. The best fit distribution for
this cluster, at the 65\% level, is achieved when the BSS were formed
at a constant rate over the entire lifetime of the cluster, right up
to the present day. This result is significantly different from the
other five clusters investigated in this paper. Perhaps, it could be
an indication of the different formation histories of binary mergers
and collisions. It is also possible that the difference between
this cluster and the other 5 could be a result of different
evolutionary tracks for binary merger and collision products.
 
\subsection{Indications and Trends} 
 
The first conclusion that we draw from the models is that since we are 
limiting our study to BSS brighter than the limit of $m_{255} = 19.0$, 
we have no information about BSS formation earlier than about 5 Gyr 
ago. On the other hand, we can state with certainty that BSS have been 
formed in all these clusters during the last 5 Gyr. In all clusters, 
this formation is not confined to the very recent past; rather, it has 
lasted over the entire time that we can probe, and in fact seems to be 
concentrated at earlier epochs. 
 
The KS tests on the luminosity and temperature functions give the best 
results for distributions in which BSS stopped forming 1--2 Gyr ago in 
all clusters except NGC 288. These ``best'' fits are clearly not good 
fits, however. In most of the clusters, there is a population of 
bright BSS that was formed more recently. Our models 
consistently under-predict the number of bright BSS compared to the 
number of fainter, redder BSS. There are a number of possible reasons. 
 
The first, and most likely, discrepancy is in the formation rate of
blue stragglers.  We have assumed a constant formation rate over the
some times, and zero otherwise. If, instead, blue stragglers were
formed at some (more or less) constant rate between 5 and 2 Gyr ago,
and then experienced a reduced (but not zero) blue straggler formation
rate to the present day, the predicted distribution would match the
observations better. It is very unlikely that the blue straggler
formation rate changes sharply between zero and a constant
value. Rather, the formation rate is going to be smoother function of
central density, velocity dispersion, single star mass function and
the number and nature of binary systems in the cluster. The simple
models used in this paper should be taken as indications of formation
eras only.
 
It is also possible that the stellar evolutionary models which go into 
these blue straggler distributions are not accurate. If we were 
overestimating the lifetimes of the faint blue stragglers or 
underestimating the lifetimes of the bright blue stragglers, we would 
see this kind of mismatch between observations and data. It is 
difficult to understand how the evolutionary tracks would be 
overestimating the lifetimes of the faint blue stragglers. The lower 
the mass of the collision product, in general, the more similar it is 
to a normal star of the same mass, and so its evolutionary track is 
very much like normal stars---something we understand quite well. In 
order to underestimate the lifetimes of the brighter blue stragglers, 
the collision products would have to mix more hydrogen into their 
cores than is currently predicted. Rapid rotation is a plausible 
mechanism for this mixing, although we would have to have some process 
by which the lower mass BSS were not rotating as rapidly. It is also 
possible that the bright blue stragglers could actually be the merger 
of {\it three} stars rather than two (as is expected to happen in 
binary-mediated collisions), and so our evolutionary models will not 
be applicable. The merger product of three $0.4\,M_{\odot}$ stars will 
have a higher percentage of hydrogen in its core than the product of 
two $0.6\,M_{\odot}$ stars, but approximately the same mass. 
 
The difference between M3 and M92 on one hand, and M10, M13 and M80 on 
the other is primarily the existence of a few very bright blue 
stragglers in the first group of clusters. In all 5 clusters, the blue 
stragglers fainter than about $m_{255}=18.5$ are very well modeled by 
the same distribution (a truncated formation scenario, suggesting that 
blue straggler formation rates should be weighted towards 3--5 Gyr 
ago). The fact that M3 and M92 are better fit by a formation scenario 
in which BSS formation ended more recently is simply a reflection of 
the number of very bright BSS in these clusters. These bright blue 
stragglers could be an indication of continued BSS formation, or of a 
different binary distribution (which produced more triple
collisions). M3 does have a significant population of BSS in the outer
part of the cluster which are thought to be the result of binary mergers
(Ferraro \etal 1997a). This fact does set M3 apart from M13, M10, \&
M80, which have few exterior BSS. Unfortunately M92 also has few
exterior BSS so they cannot be invoked to explain the M3/M92 similarity.
 
As noted in Section 3.1, another very interesting point is that M10,
M13 and M80 (which share a common BSS Luminosity function) all have
blue tail components to their horizontal branches, while M3, M92 and
NGC 288 do not. M3 and M13 are also a classic second-parameter pair
(see Ferraro \etal 1997c),
referring to the difference in their horizontal branch morphology
without an obvious difference in any other cluster parameter: they
have the same metallicity, central density, total mass, and very
similar velocity dispersions. (The two clusters may have slightly
different ages, which may or may not affect horizontal branch
morphology---see Rey \etal 2001, Davidge \& Courteau 1999, Ferraro
\etal 1997c). From these data, there is certainly a suggestion that
BSS population could be connected to the second parameter. From the
comparisons in this paper, it seems that it is not the total number of
BSS or their specific frequency, but rather the kind of BSS which are
produced in ``blue tail'' clusters which is different from those
produced in non-blue tail clusters. If this link between BSS and
horizontal branch morphology is real, then it suggests that the second
parameter could be related to either stellar collisions and
encounters, or to binary evolution. 
 
\section{Conclusions} 
\label{sec:con} 
 
In this paper, we have compared the blue straggler distribution in six
galactic globular clusters: M3, M10, M13, M80, M92 and NGC 288. The
blue stragglers were observed in the HST UV filters F255W and
F336W. The clusters have similar total masses, and velocity
dispersions; have intermediate to low metallicities and span a range
in central density. We studied the properties of the blue stragglers
relative to other cluster populations; and we compared the observed
blue straggler distributions with theoretical models of collision
products in globular clusters.
 
Whether they formed via a
collisional channel or a merged primoridial binary channel, BSS are the
most massive stars in GGCs other than neutron stars or possibly some white
dwarfs. Since dynamical relaxation times are typically shorter than
BSS lifetimes they should settle toward the cluster center. BSS formed
from merged primordial binaries have the entire cluster lifetime to
settle. Collisions occur most often in high density regions and most
often involve binaries providing yet another reason to expect BSS to
be found near cluster centers. Indeed, it is difficult to image a BSS
formation scenario which would not lead to a centrally condensed BSS
distribution. Our results are consistent with this expectation: M3,
M92, \& M80 show strong central concentrations; the BSS in M10 and
NGC~288 are most likely centrally condensed; while the BSS in M13
could have the same distribution as the RGB, they could also have a
central concentration similar to that of M3. The small sample size is
not adequate to definitively determine the radial distribution. 
 
The specific frequency of blue stragglers compared to the number of 
horizontal branch stars varies from 0.07 to 0.92 for these six 
clusters, and does not seem to be correlated with central density, 
total mass, velocity dispersion, or any other obvious cluster 
property. We do not have measurements of the binary fraction in most 
of these clusters. Binary stars impact the blue straggler populations 
in a number of ways. BSS can be formed from binary evolution in any 
environment, although binary systems in dense environments will be 
different from those in sparse environments---close binaries will be 
hardened by encounters (increasing the number of ``binary merger'' 
BSS), but wide binaries are also disrupted (decreasing the number of 
``binary merger'' BSS).  The relative rates of these events is an 
important factor in looking at blue stragglers in clusters. Binary 
systems also facilitate close interactions between stars because they have a 
large cross section compared to single stars, so we expect that 
clusters with a large binary fraction to have both more collisional
and binary merger blue stragglers. 
 
Fusi Pecci \etal (1992, and references therein) presented a 
qualitative interpretative scenario concerning the possible origin of 
BSS in GGCs. They suggest that BSS in loose clusters might be produced 
from coalescence of primordial binaries. In high density GGCs 
(depending on survival-destruction rate of primordial binaries) BSS 
might mostly arise from stellar interactions, particularly those which 
involve binaries (Ferraro, Fusi Pecci \& Bellazzini, 1995). 
 In this scenario one could find clusters where more 
than one mechanism is at work in generating BSS.  In fact Ferraro 
\etal (1993, 1997a) found that the radial distribution and the 
luminosity functions of BSS in the center of M3 are consistent with a 
collisional origin, while in the outer regions they are consistent 
with a primordial binary origin.  The paucity of BSS in M13 suggests that 
either the primordial population of binaries in M13 was poor or that 
most of them were destroyed.  Alternatively, as suggested by Ferraro \etal 
(1997b), the mechanism producing BSS in the central region of M3 is 
more efficient than M13 because M3 and M13 are in different dynamical 
evolutionary phases. 
 
The evidence shown above suggests that different channels of BSS
formation should be at work and/or a large difference in the binary
fraction should be present in NGC 288 in order to produce such a large
BSS frequency in a cluster with such different structural
parameters. Bellazzini \etal (2002) measured the binary fraction in
the core of NGC 288 to be between 10\% and 38\%. Similar measurements
for M3, M13, and M80 will be extremely helpful in solving this
mystery.
 
According to simple models, BSS formation occured in all clusters over 
last 5 Gyr at least, and more were formed 2--5 Gyr ago than in the 
recent past. This result needs to be tested using models of globular 
cluster evolution in which the feedback between stellar collisions and 
cluster evolution is modeled explicitly. Our assumption of a BSS 
formation rate which is either constant or zero is unphysical, and 
more complicated models are clearly required. 
 
Finally, we note the possible connection between horizontal branch
morphology and blue straggler population characteristics. Of our six
clusters, three have HB blue tails (M10, M13, M80), and three do
not. The three with blue tails have the same blue straggler luminosity
function, which spans about 1.5 magnitudes compared to 2.5 magnitudes
in M3 and M92.  This suggests that blue straggler populations and the
horizontal branch second parameter may be linked in some way, perhaps
through the complicated world of binary systems and their effect on
cluster evolution and populations.

\acknowledgments

We warmly thank Michele Bellazzini for useful discussions and the
 referee, Peter Leonard, for the many helpful suggestions which
 significantly improved the presentation of the paper.  The financial
 support of the Agenzia Spaziale Italiana (ASI) and of the {\it
 Ministero dell'Istruzione, dell'Universit\`a e della Ricerca} (MIUR)
 is kindly acknowledged. RTR is partially supported by grant GO-8709
 from STScI.


\newpage

\begin{figure} 
\plotone{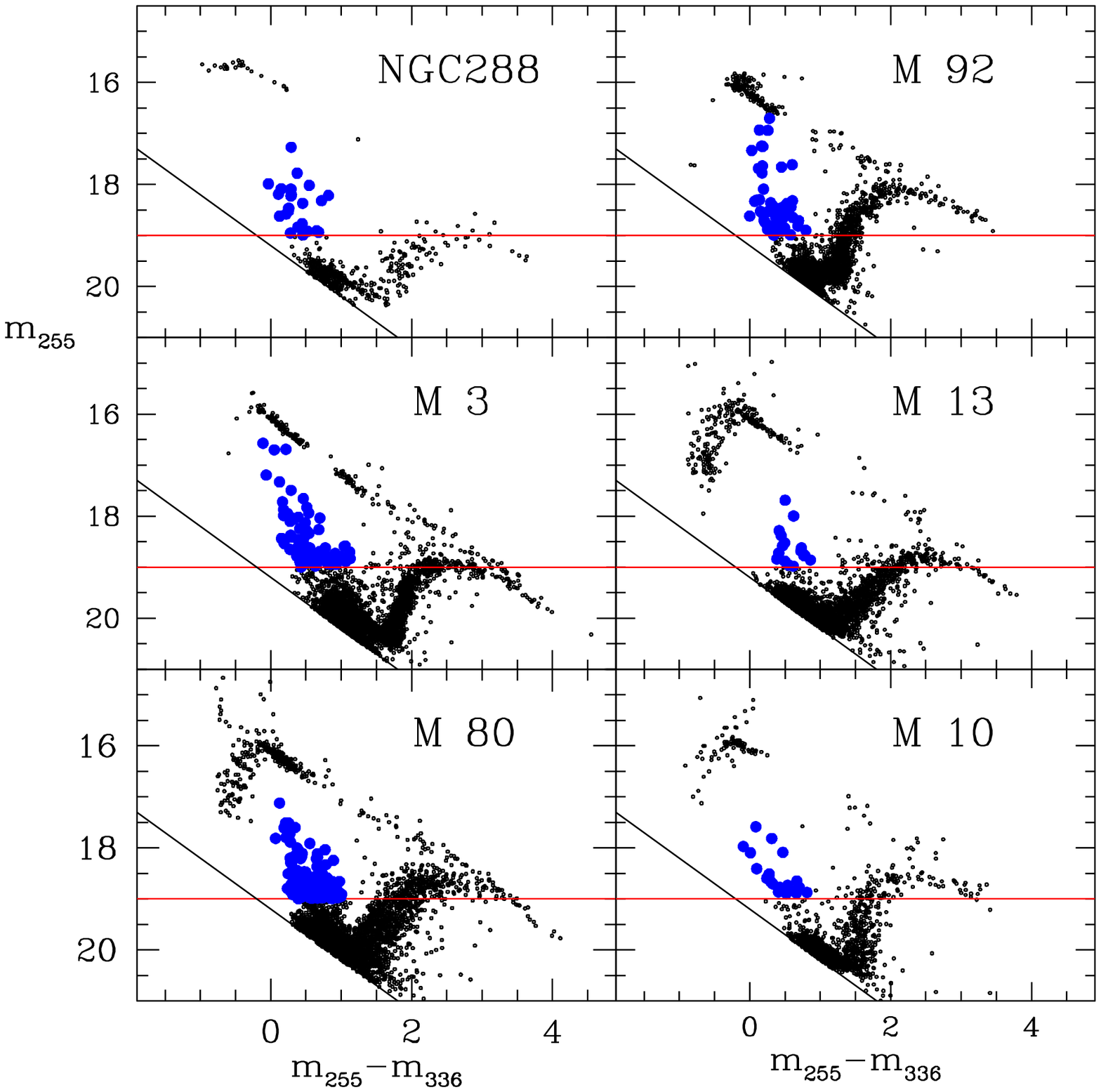} 
\caption{($m_{255}, m_{255}-m_{336}$) CMDs for the selected clusters. 
Horizontal and vertical shifts have been applied to all   CMDs in order  
to match the main sequences of M3.  
The horizontal solid line corresponds to $m_{255}=19$ in M3. 
The bright BSS candidates are marked as large empty circles.  
\label{fig:map336}} 
\end{figure} 
 
\begin{figure} 
\plotone{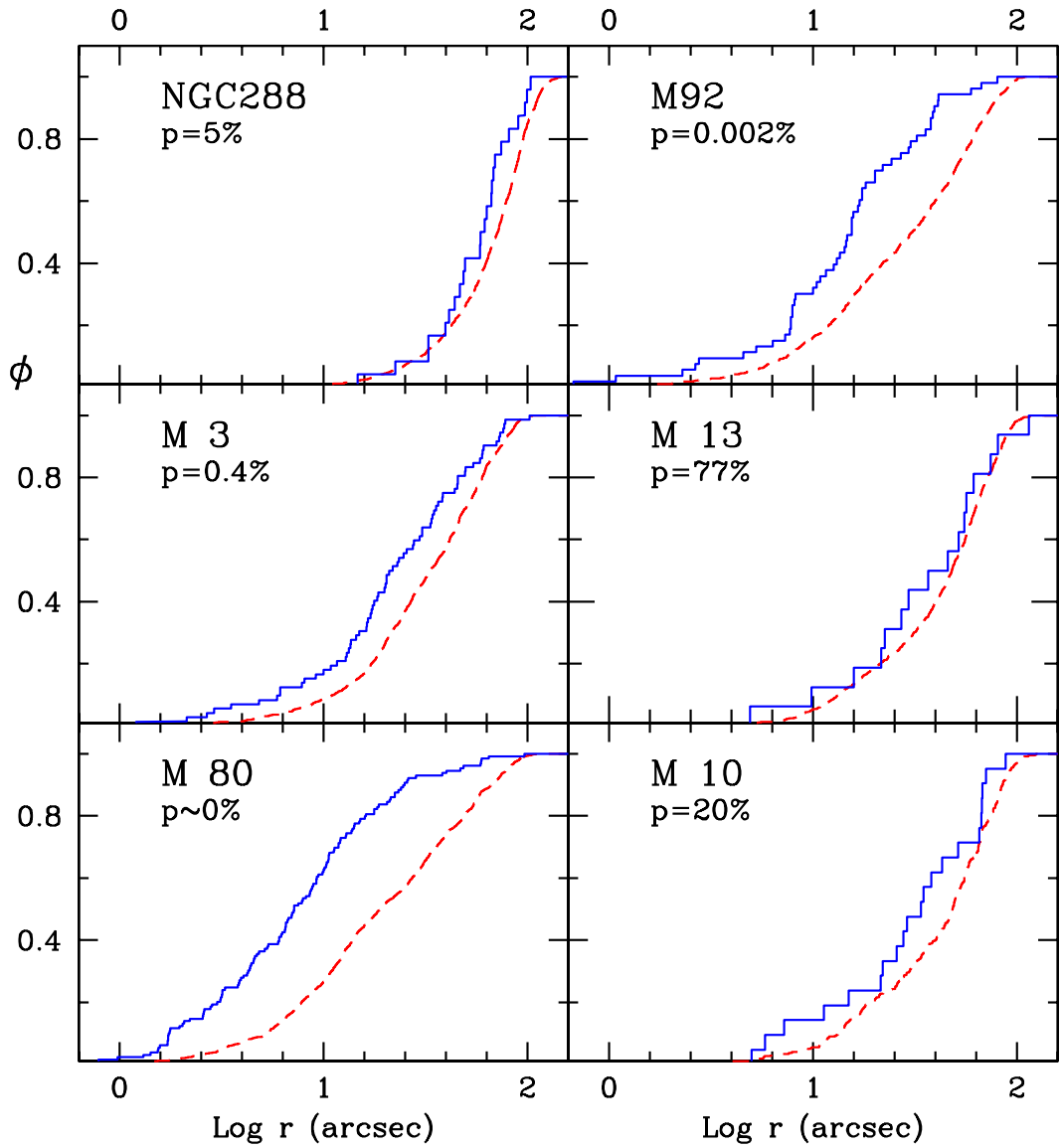} 
\caption{ 
Cumulative radial distributions for the {\em bright} 
 BSS (solid line) with respect to the RGB stars (dashed line) as a  
function of their projected distance  
($r$) 
from the cluster center for each of the  
six clusters. 
The probability that the two  populations are extracted from the 
same  distribution is also reported in each panel. 
\label{fig:raddist}} 
\end{figure}

\begin{figure}
\plotone{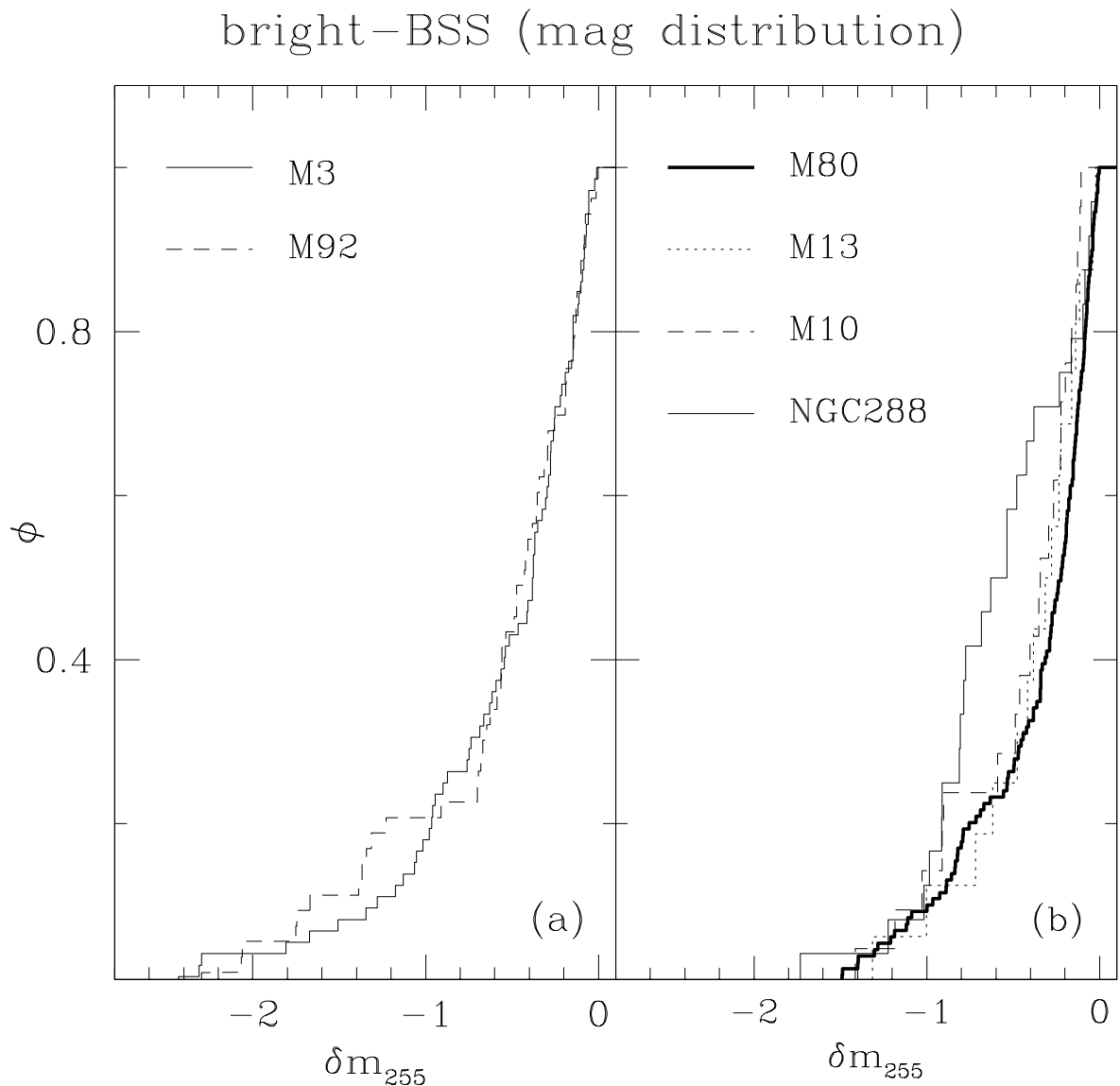} 
\caption{ 
Cumulative  magnitude distributions for the {\it bright} 
 BSS   for each of the  
 six clusters. The $\delta m_{255}$ parameter is the 
 difference in magnitude with respect to the limit threashold 
  ($m_{255}=19$) for bright-BSS, see text. 
 In {\it Panel (a)} the BSS distributions 
 for  M3 and M92 (the two clusters for which the  
 BSS distribution extends up to more than  
 two magnitudes brighter than the threashold) are compared. 
 In  {\it Panel (b)} the BSS magnitude distributions for the other 4 
  clusters are plotted.   
\label{fig:magdist}} 
\end{figure} 

\begin{figure}
\plotone{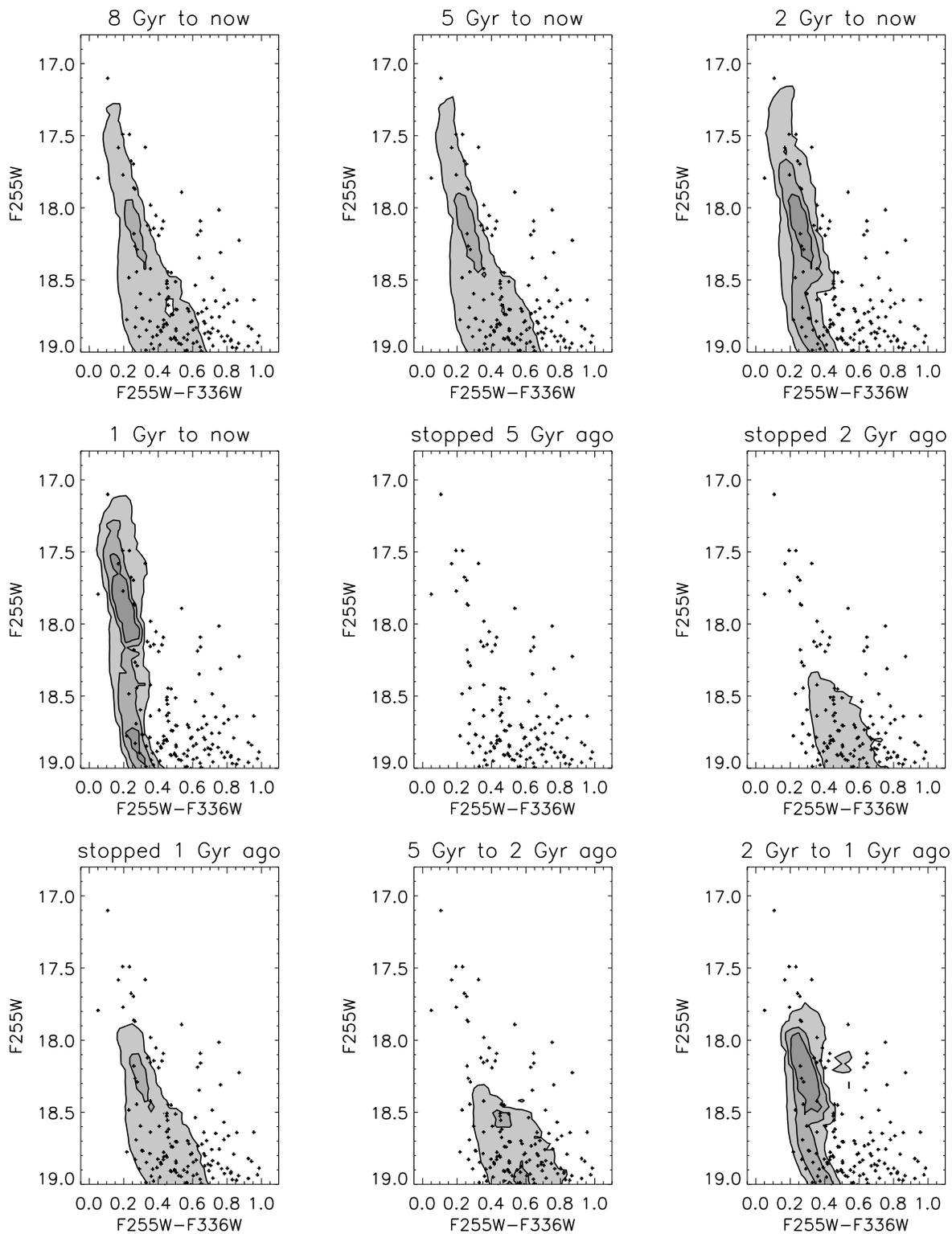} 
\caption[f4.ps]{The distribution of bright blue stragglers 
in the color-magnitude diagram for M80, compared to theoretical 
distributions. The observations are plotted as crosses, and the 
greyscale contours give the theoretical distributions, with darker 
colors indicating more blue stragglers. The different panels 
correspond to different eras of constant blue straggler formation, as 
indicated at the top of each panel. \label{fig:distributions}} 
\end{figure} 
  
\begin{figure}
\plotone{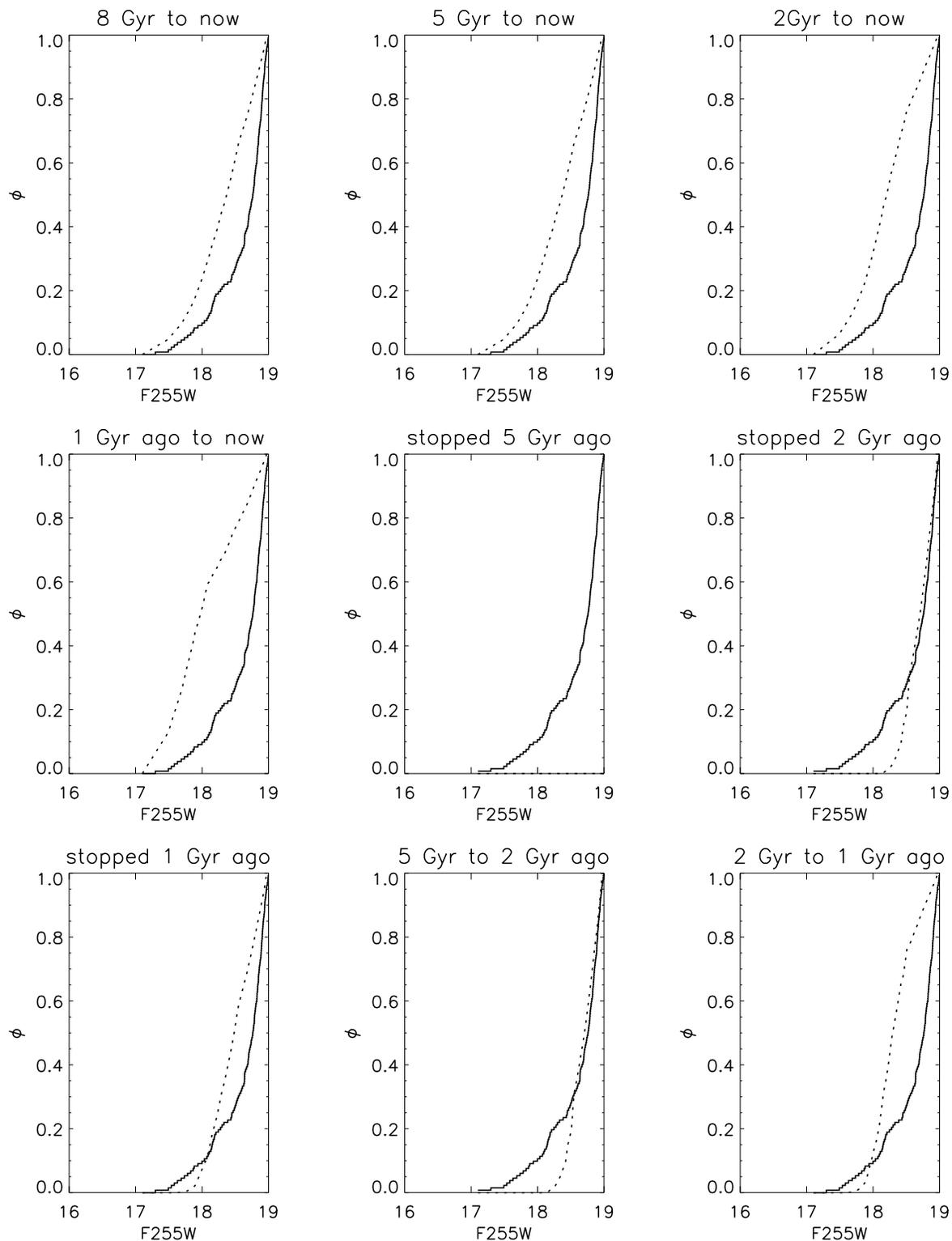}
\caption[f5.ps]{Luminosity functions for the bright blue stragglers in  
M80 (solid line), compared to theoretical predictions (dotted 
line). The different panels correspond to different eras of constant 
blue straggler formation, as indicated at the top of each panel. The 
luminosity functions are good probes of the mass function of blue 
stragglers. This figure shows that the theoretical models do not 
accurately model the relative numbers of fainter and brighter blue 
stragglers.\label{fig:lumfunc}} 
\end{figure} 
 
\begin{figure} 
\plotone{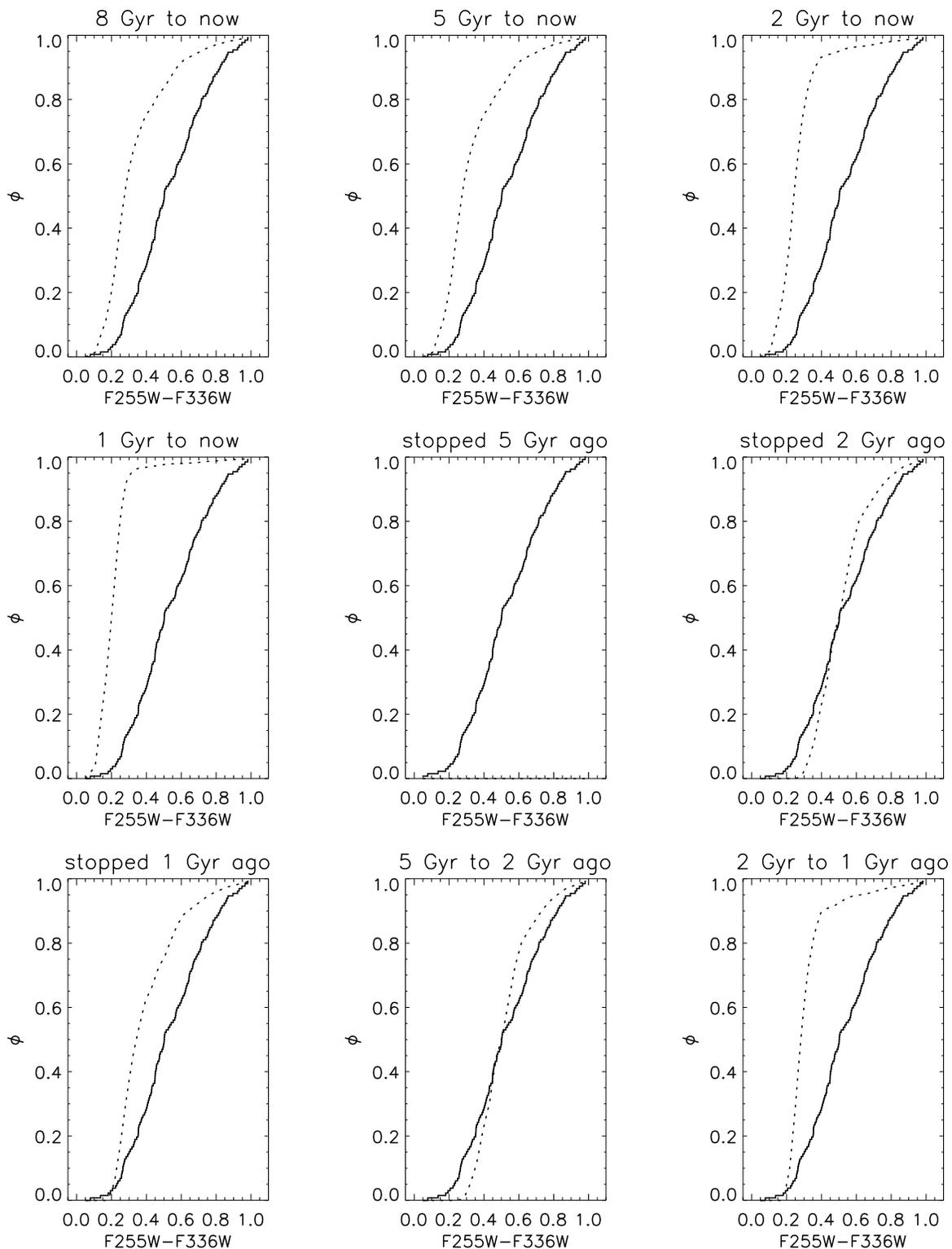} 
\caption[f6.ps]{Temperature functions for the bright blue stragglers in  
M80 (solid line), compared to theoretical predictions (dotted 
line). The different panels correspond to different eras of constant 
blue straggler formation, as indicated at the top of each panel. The 
temperature functions are good probes of the bulk evolutionary state 
of the population of blue straggler stars, indicating whether the 
stars are all on the main sequence (as in the panel marked ``1 Gyr to 
now'') or closer to the subgiant \& giant branches (as in the panel 
marked ``stopped 2 Gyr ago''). The BSS in M80 show a significant 
spread in evolutionary state. \label{fig:tempfunc}} 
\end{figure} 
 
\begin{figure}
\plotone{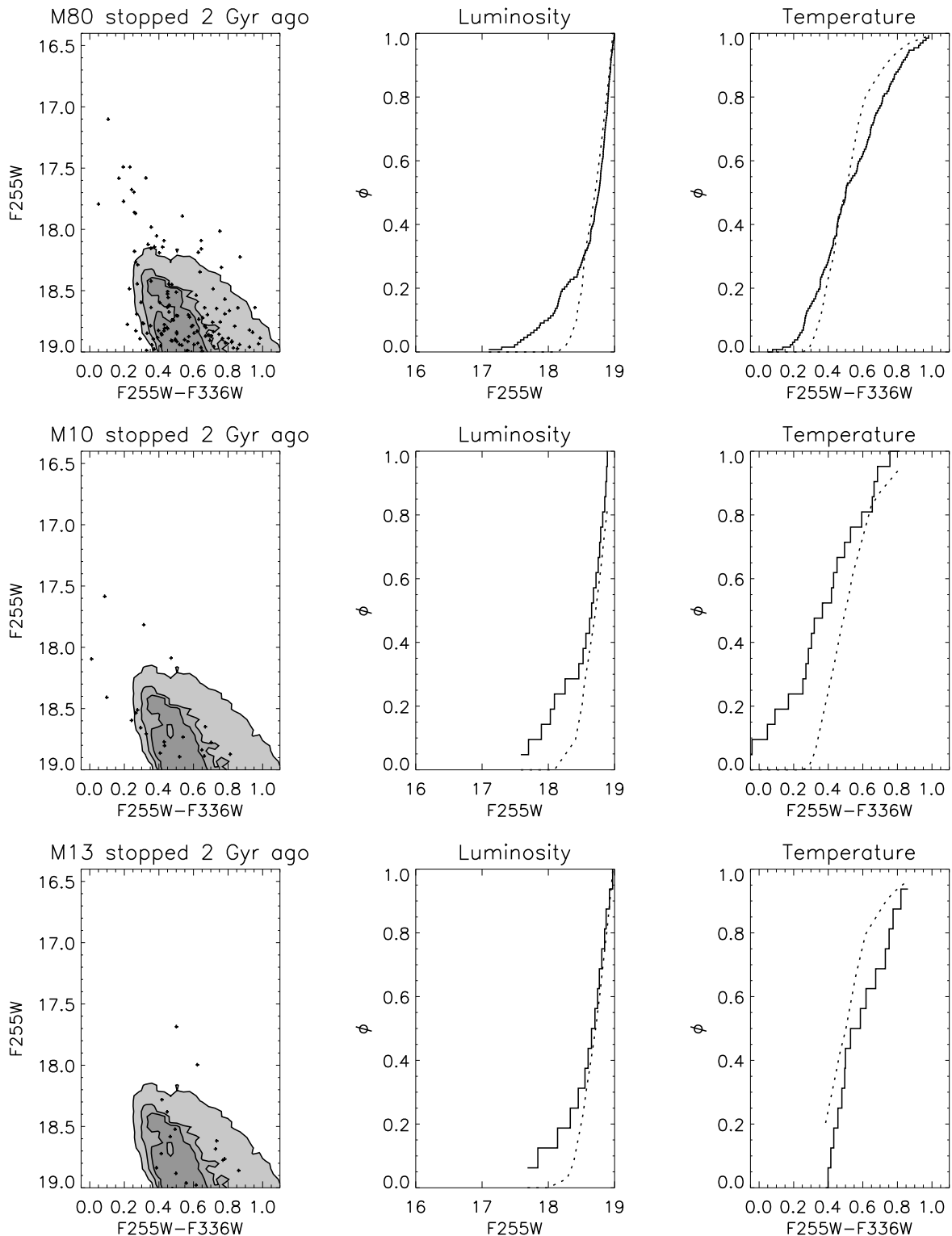}  
\caption[f7a.ps]{The observed blue stragglers (crosses in the CMD,  
solid lines in the luminosity and temperature functions) and the best 
fit theoretical distribution, as determined from KS tests on the 
luminosity and temperature functions. This figure shows the results 
for M80, M10 and M13. In all cases, the nominal best-fit distribution 
is one in which the blue stragglers stopped forming 2 Gyr 
ago. However, only for M13 is this actually plausible. M80, and to a 
lesser extent M10 were forming blue stragglers more recently than 2 
Gyr ago, since we see a population of BSS brighter than 
$m_{255}=18$. On the other hand, the temperature function (i.e. the 
distribution of BSS in evolutionary state) is very well fit by this 
truncated distribution. \label{fig:clustersA}} 
\end{figure} 
 
\begin{figure} 
\plotone{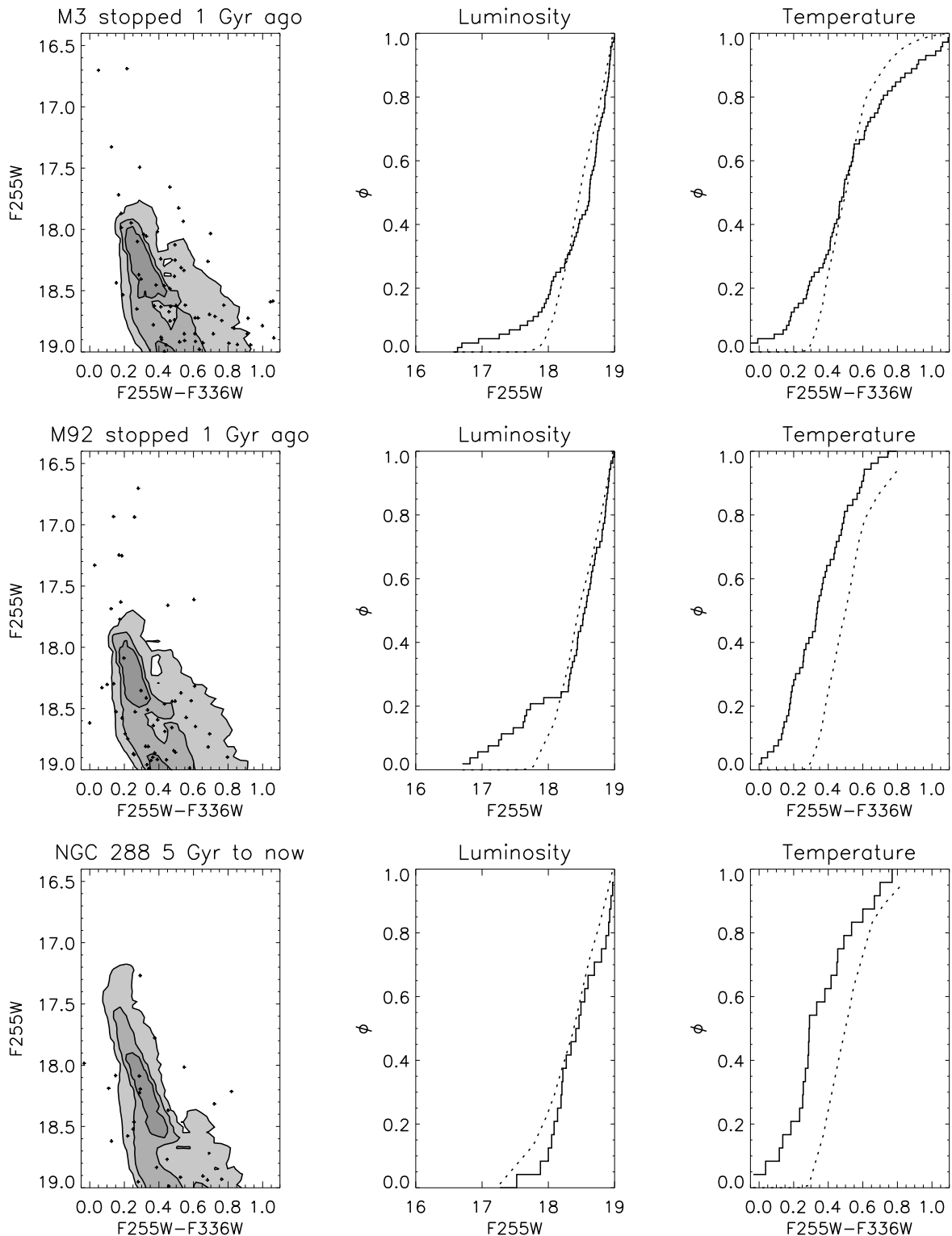}
\caption{Same as figure \ref{fig:clustersA}, for M3, M92 and NGC 288.  
M3 and M92 contain the brightest blue stragglers in our sample, and so 
the best-fit theoretical distribution is one in which young blue 
stragglers (up to 1 Gyr old) are found. However, because of the spread 
in temperature, these clusters are best fit by a truncated blue 
straggler formation model, like M80, M10 and M13. NGC 288 is different 
from the previous 5 clusters---its blue straggler distribution is 
extremely well fit by a model which has had essentially constant blue 
straggler formation over the lifetime of the cluster right to the 
present day.\label{fig:clustersB}} 
\end{figure}

\newpage

\begin{deluxetable}{cccccc} 
\scriptsize 
\tablewidth{13cm} 
\label{tab:uve} 
\tablecaption{ Cluster parameters} 
\startdata \\
\hline\hline
{Cluster} & 
{$[{\rm Fe/H]}$} & 
{$Log\rho_0$} & 
{$Mass$} &   
{$d$} &  
{$\sigma_0$}\\
 & & \small{$[M_{\odot}\,pc^{-3}]$}
 & \small{$[Log(M/M_{\odot})]$}
 &\small{$[{\rm kpc}]$}
 &\small{$[{\rm km\,s^{-1}}]$}
 \\
 \hline
     NGC 5272(M3)   & $-1.66$ & 3.5 & 5.8 & 10.1 & 5.6\\ 
     NGC 6205(M13)  & $-1.65$ & 3.4 & 5.8 &  7.7 & 7.1\\ 
     NGC 6093(M80)  & $-1.64$ & 5.4 & 6.0 &  9.8 &12.4\\ 
     NGC 6254(M10)  & $-1.60$ & 3.8 & 5.4 &  4.7 & 5.6\\ 
     NGC 288        & $-1.40$ & 2.1 & 4.9 &  8.8 & 2.9 \\ 
     NGC6341 (M92)  & $-2.24$ & 4.4 & 5.3 &  9.0 & 5.9\\ 
\enddata 
\end{deluxetable} 
 
\begin{deluxetable}{ccc} 
\scriptsize 
\tablewidth{9cm} 
\label{tab:exptime} 
\tablecaption{WFPC2 exposure times} 
\tablehead{ 
\colhead{Cluster} & 
\colhead{F336W-Exp [s]} &  
\colhead{F255W-Exp [s]}  
} 
\startdata 
       M3   &   3340  & 1200  \\ 
      M13  &   560  & 200   \\ 
      M80  &   2400  & 1160 \\ 
      M10  &   1500  & 5200 \\ 
     NGC 288 & 3760  & 700 \\ 
      M92    & 3800  & 2200 \\ 
\enddata 
\end{deluxetable}

\newpage
\begin{deluxetable}{ccccccccc} 
\scriptsize 
\tablewidth{14cm} 
\label{tab:populations} 
\tablecaption{Bright BSS populations and distributions  } 
\tablehead{ 
\colhead{Cluster} & 
\colhead{$L_S/L_T$} &
\colhead{$N_{\scriptscriptstyle\rm bBSS}$} &
\colhead{$S4_{\scriptscriptstyle\rm BSS}$} & 
\colhead{\bbsshb} & 
\colhead{ \rhalf  } & 
\colhead{ $r_c$ } & 
\colhead{\rhalf/\rcore} &  
\colhead{\frgb} 
} 
\startdata 
       M3 & 0.25 & 72 & 6.0 &0.28 & 22\arcsec & 30\arcsec &0.73 & 0.31 \\ 
      M13 & 0.27 & 16 & 1.6 &0.07 &46\arcsec & 40\arcsec  &1.15 & 0.47 \\ 
      M80 & 0.48 &129 & 13.4 &0.44 & 7\arcsec  & 6.5\arcsec  &1.07 & 0.17 \\ 
      M10 & 0.30 & 22 & 5.6 &0.27 & 34\arcsec & 40\arcsec &0.85 & 0.35 \\ 
   NGC 288& 0.20 & 24 & 21.8&0.92 &  60\arcsec & 85\arcsec & 0.71 & 0.38 \\ 
     M92  & 0.45 & 53 & 4.8 & 0.33 &  15\arcsec & 14\arcsec & 1.07 & 0.29 \\ 
\enddata 
\end{deluxetable}

\begin{deluxetable}{cccc} 
\scriptsize 
\tablewidth{10cm} 
\label{tab:bb} 
\tablecaption{ Expected number of binary-binary encounters per Gyr } 
\tablehead{ 
\colhead{Cluster} & 
\colhead{$a_{hs} (AU)$}& 
\colhead{$N_{\scriptscriptstyle\rm bb}(f_{b}=0.2)$}& 
\colhead{$N_{\scriptscriptstyle\rm bb}(f_{b}=1)$}   
} 
\startdata 
      M3   &  0.63 &   12 &   304 \\ 
      M13  &  0.39 &   4 &  99 \\ 
      M80  &  0.13 &   66 &  1640 \\ 
      M10  &  0.63 &   12 & 289 \\ 
    NGC 288&  2.36 &   2 &  52 \\ 
     M92  &   0.57 &   47 &   1178 \\ 
\enddata 
\end{deluxetable}

\end{document}